\documentclass[superscriptaddress,twocolumn,english,pra,aps]{revtex4}
\usepackage[T1]{fontenc}
\usepackage[latin9]{inputenc}
\usepackage{amsmath}
\usepackage{amssymb}
\usepackage{graphicx,CJK}
\usepackage{epstopdf}
\usepackage{color}
\usepackage{hyperref}
\usepackage{gensymb}

\makeatletter

\usepackage{mathrsfs}

\makeatother

\usepackage{babel}

\begin{document}
%\begin{CJK*}{GBK}{song}

\title{Optical nonreciprocity in rotating diamond with nitrogen-vacancy center}% Force line breaks with \\

%\author{Hong-Bo Huang(»Æãü²©),$^{}$ Jun-Jie Lin(ÁÖ¿¡œÜ),$^{}$  Yi-Xuan Yao(ÒŠÒÀÝæ),$^{}$
%and Qing Ai(°¬Çå)$^{}$\footnote{aiqing@bnu.edu.cn\homepage{http://quanphys.bnu.edu.cn}}}
%
\author{Hong-Bo Huang}

\address{Department of Physics, Applied Optics Beijing Area Major Laboratory,
Beijing Normal University, Beijing 100875, China}

\author{Jun-Jie Lin}

\address{Department of Physics, Applied Optics Beijing Area Major Laboratory,
Beijing Normal University, Beijing 100875, China}

\author{Yi-Xuan Yao}

\address{Department of Physics, Applied Optics Beijing Area Major Laboratory,
Beijing Normal University, Beijing 100875, China}

\author{Ke-Yu Xia}

\address{National Laboratory of Solid State Microstructures,
Collaborative Innovation Center of Advanced Microstructures, College of Engineering and Applied Sciences, and School of Physics, Nanjing University, Nanjing 210093, China}

\author{Zhang-Qi Yin}

\address{Center for Quantum Technology Research and Key Laboratory of
Advanced Optoelectronic Quantum Architecture and Measurements
(MOE), School of Physics
Beijing Institute of Technology
Beijing 100081, China}

\author{Qing Ai}
\email{aiqing@bnu.edu.cn}
\address{Department of Physics, Applied Optics Beijing Area Major Laboratory,
Beijing Normal University, Beijing 100875, China}

%\author{Hong-Bo Huang,$^{1}$ Jun-Jie Lin,$^{1}$  Yi-Xuan Yao,$^{1}$
%and Qing Ai\footnote{aiqing@bnu.edu.cn}$^{1}$}
%
%\address{Department of Physics, Applied Optics Beijing Area Major Laboratory,
%Beijing Normal University, Beijing 100875, China\\}

\date{\today}% It is always \today, today,

\begin{abstract}
We theoretically propose a method to realize optical nonreciprocity in rotating nano-diamond with a nitrogen-vacancy (NV) center. Because of the relative motion of the NV center with respect to the propagating fields, the frequencies of the fields are shifted due to the Doppler effect. When the control and probe fields are incident to the NV center from the same direction, the two-photon resonance still holds as the Doppler shifts of the two fields are the same. Thus, due to the electromagnetically-induced transparency (EIT), the probe light can pass through the NV center nearly without absorption. However, when the two fields propagate in opposite directions, the probe light can not effectively pass through the NV center as a result of the breakdown of two-photon resonance.
%Beyond the optical isolation, our proposal provides an alternative avenue for measuring the angular velocity of rotation and realizing an effective gyroscope.

%Recent studies have shown that non-reciprocal quantum systems can be induced by random
%thermal motions [S. C. Zhang, \textit{et al}., Nat. Photon. \textbf{12}, 744 (2018)]. The direct reason for this nonreciprocity is the Doppler shift caused by the relative motion between atoms and light. Here, we demonstrate a scheme of non-reciprocal transmission by using a high-speed rotating diamond and explore the influence of NV color center rotation on nonreciprocity. By measuring the isolation degree of the system, we can measure the rotating speed of the diamond and calculate the measurement sensitivity. Thus, our works provide a comprehensive way of understanding the Doppler-shift-induced nonreciprocal quantum systems and can be used in the manufacture of gyroscope which may inspire further studies and possible applications.
\end{abstract}
\maketitle
%\end{CJK*}

\section{\label{sec:Intro}Introduction}

Optical nonreciprocity happens when Lorentz's reciprocity is broken, and it leads to diffirent transmittance when two beams of light propagate in the opposite directions \cite{Born1999}. Optical nonreciprocity plays an important role in optical devices such as isolators and circulators \cite{Jalas2013}, which have further application in quantum networks \cite{Kimble2008,Lodahl2017}, quantum noise reduction \cite{Shoji2008}, quantum signal processing \cite{Khanikaev2015,Yu2009}, and photon blockade \cite{Maayani2009,Huang2018,Huang2012}. Traditional nonreciprocity is mainly realized by magneto-optical effect \cite{Tzuang2014,Khanikaev2010,Bi2011}, which often requires such large size that it is difficult to be used on chips. To overcome this shortcoming, alternative strategies are explored to achieve nonreciprocity, including nonlinear optics \cite{Bender2013,Fan2012,Chang2014}, synthetic magnetism \cite{Fang2012,Fang2012-2,Tzuang2014}, optomechanical coupling \cite{Ruesink2016,Xu2015,XWXu2018}, and non-trivial topology \cite{Zhu2021}. Interestingly, it has been shown that nonreciprocal transport can also be realized by the irregular thermal motion of atoms \cite{Liang2020,Dong2021,Zhang2018,Lin2019}. %Therefore, the exploration of nonreciprocity is of great significance.

In 1961, Fano pointed out that if several different atomic transitions are coupled, the total transition probability will be coherently enhanced or cancelled due to interference of the amplitudes of these transitions \cite{Fano1961}. Inspired by this discovery, many studies on atomic coherence appeared, e.g. electromagnetically induced transparency (EIT) \cite{Harris1990}. In EIT, adding a strong control field can make the probe light transparent, which will be resonantly absorbed by the atomic medium in the absence of the control field \cite{Scully1997}. Because it has a wide range of prospective applications, e.g. nonreciprocal transmission and memory of nonclassical fields \cite{Lobino2009,Honda2008}, EIT has been successfully realized in many systems, such as gas-phase atoms \cite{Akulshin1998,Figueroa2010}, photosynthetic energy transfer \cite{Dong2012,LXu2018}, metamaterial \cite{Papasimakis2008}, superconducting system \cite{Anisimov2011}, and NV center in diamond \cite{WangYY2018}.

On the other hand, nitrogen-vacancy (NV) center in diamond is an intriguing platform for quantum information processing \cite{ZhangHJ2021,Doherty2013,Tao2015} and quantum sensing \cite{Schirhagl2014,Li2017}. NV center is a pair of point defects at adjacent sites in diamond crystal. Due to long coherence time and easy manipulation at room temperature, it has been shown that it can be utilized for detecting magnetic cluster \cite{Zhao2011-1}, state transfer by shortcut to adiabaticity \cite{Zhou2017,Song2016-1,Song2016-2}, gyroscope \cite{Ledbetter2012,Ajoy2012}, and quantum hyperbolic metamaterial \cite{Ai2021}. Recently, it has been experimentally realized that nanoparticles levitated in vacuum can rotate at a frequency of GHz \cite{Reimann2018,Ahn2018,Hoang2016}. Inspired by the rapid progress on the quantum coherent devices by NV centers, a question naturally comes to our mind: Can we make use of a rotating diamond with NV centers for realizing optical nonreciprocity?

In this paper, we propose a non-reciprocal transmission based on a rotating nano-diamond at a high speed. The nano-diamond doped with NV centers is placed in an optical cavity. Two electronic ground states and one electronic excited state of the NV center form a $\Lambda$-type three-level configuration and the optical transitions between the ground and excited states can be induced by electromagnetic fields, i.e., the control field and cavity field. We explore the transmittance of the probe light incident in different directions with respect to the control field.

The paper is organized as follows: In Sec.~\ref{sec:Model}, we first introduce the model of the system and we derive the expressions of the transmittance at the steady state by the Heisenberg-Langevin approach. In Sec.~\ref{sec:Results}, we numerically show the transmittance for the two cases and analyze the results by the dark-state mechanism in Ref.~\cite{WangYY2018}. Finally, we discuss the prospect of our proposal and summarize the main findings in Sec.~\ref{sec:Conclusion}.

\section{Theoretical Model}
\label{sec:Model}

\begin{figure}
\includegraphics[scale=0.22]{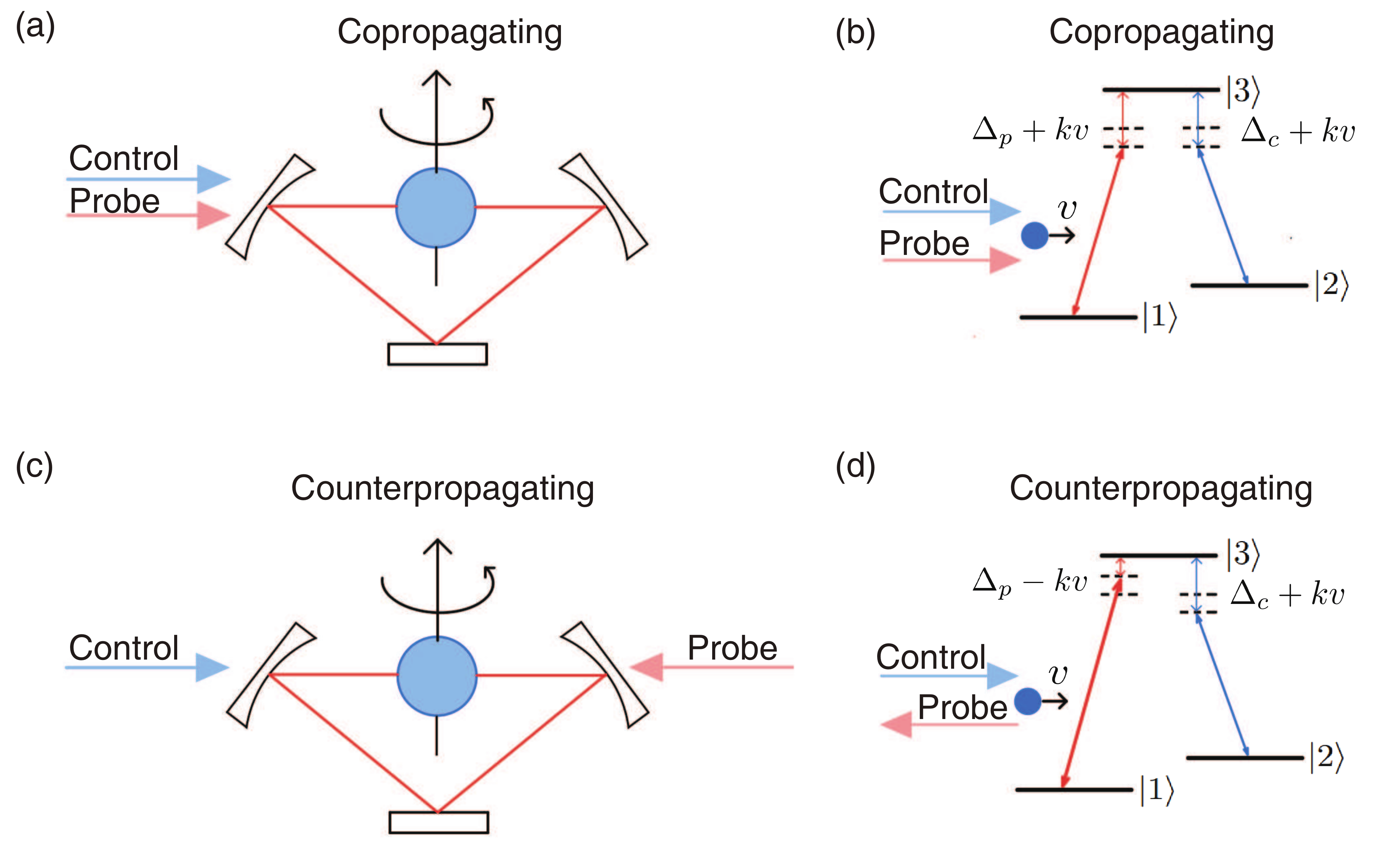}
\caption{Schematic diagram of optical nonreciprocity in a rotating nano-diamond with an NV center. (a) and (c) are the cases when the probe and the control field propagate in the same/opposite direction, respectively. A high-speed rotating nano-diamond is placed in an optical cavity. After the lights entering the cavity, they are coupled with the NV center in the nano-diamond. (b) and (d) are energy-level diagrams for (a) and (c), respectively. Due to the relative motion between the NV center and the lights, different Doppler shifts are induced. When the laser and NV center move along the same direction, the Doppler shift is $\Delta+kv$. Otherwise, the Doppler shift is $\Delta-kv$.
\label{Fig1}}
\end{figure}

We consider a rapidly-rotating nano-diamond with an NV center in an optical cavity, as illustrated in Fig.~\ref{Fig1}. A beam of control light and probe light respectively interact with the optical transitions at $\lambda=637.2$~nm of an NV center from the electronic ground state $\vert\pm1\rangle$ to the electronic excited state $\vert A_2\rangle$ \cite{Zhou2017,Ai2021}. The cavity mode with resonance frequency $\omega_{a}$ is coupled to the transition $\vert1\rangle(\equiv\vert+1\rangle)\leftrightarrow\vert 3\rangle(\equiv\vert A_2\rangle)$ with Rabi frequency $g$. Moreover, there is a probe laser injected into the cavity. A strong control laser beam with
a carrier frequency $\omega_c$ and Rabi frequency $\Omega_c$ is coupled to the transition $\vert2\rangle(\equiv\vert-1\rangle)\leftrightarrow\vert3\rangle$. %Also, we have to consider the interaction between the system and the surrounding environment. We use a simple model for the environment (reservoir), that is, it is composed of a large number of harmonic oscillators, whose frequencies have a broad spectrum distribution.

The relative motion between NV center and photons gives rise to the microscopic Doppler shift, i.e., $\omega_\alpha-\vec{k}_\alpha\cdot\vec{v}$. Here, $\omega_\alpha$ ($\alpha=a,c$) is the frequency of the photon experienced by the NV center when the latter is at rest. $\vec{k}_\alpha$ is the wave vector of the light in the nano-diamond and $\vec{v}$ is the velocity of the NV center. For example, if the NV center and light move in the same direction, the frequency of the photon experienced by the NV center becomes $\omega-kv$.

Assuming $\hbar=1$, the Hamiltonian of the system takes the following form
\begin{eqnarray}
H&=&\omega_{a}a^{\dagger}a+\omega_{p}a_p^{\dagger}a_p+{\sum^{3}_{l=1}}\omega_{l}\sigma_{ll}+\sum_{r}\omega_rd_r^\dagger d_r\nonumber\\
&&+\sum_{q}\omega_qb_q^\dagger b_q+\Omega_{c}e^{i(\omega_{c}-\vec{k}_c\cdot\vec{v})t}\sigma_{23} +ge^{-i\vec{k}_a\cdot\vec{v}t}a^{\dagger}\sigma_{13} \nonumber\\
&& +i\sqrt{\kappa_{1}}a^{\dagger}a_{p}+\sum_{r}g_rd_r^\dagger a+\sum_{q}(g_{1q}\sigma_{31}+g_{2q}\sigma_{32})b_q\nonumber\\
&&+\textrm{h.c.},
\end{eqnarray}
where $\omega_{l}$ $(l=1,2,3)$ is the energy of state $\vert{l}\rangle$, and $\hat{\sigma}_{lm}=\vert{l}\rangle\langle{m}\vert$ is the operator of NV center, $a$ and $a_{p}$ are respectively the annihilation operator of the cavity mode and the probe field, $\Omega_c$ is the Rabi frequency of the control field with frequency $\omega_c$ and wave vector $\vec{k}_c$, $g$ is the Rabi frequency of the cavity mode with frequency $\omega_a$ and wave vector $\vec{k}_a$. $b_{q}$ and $d_{r}$ are annihilation operator of the reservoir interacting with the NV center and cavity respectively, where $\omega_{q}$ and $\omega_{r}$ are the frequency of the corresponding harmonic oscillator. $g_{r}$, $g_{1q}$, and $g_{2q}$ correspond to the coupling strength between the reservoir and the cavity, atomic transition $\vert1\rangle\leftrightarrow\vert3\rangle$, and $\vert2\rangle\leftrightarrow\vert3\rangle$, respectively.

By using the unitary transformation
\begin{align}
U=&\exp\{i[\omega_{p}a^{\dagger}a+\omega_{p}a_{p}^{\dagger}a_{p}+\sum_{r}\omega_{r}d_{r}^{\dagger}d_{r}\nonumber\\
&+\sum_q\omega_{q}b_{q}^{\dagger}b_{q}+(\omega_{3}-\omega_{p}+\vec{k}_a\cdot\vec{v})\sigma_{11}\nonumber\\
&+(\omega_{3}-\omega_{c}+\vec{k}_c\cdot\vec{v})\sigma_{22}+\omega_{3}\sigma_{33}]t\},
\end{align}
we can obtain the Hamiltonian in the interaction picture as
\begin{align}
H_I&=\Delta_{a}a^{\dagger}a+(\Delta_{p}+\vec{k}_a\cdot\vec{v})\sigma_{11}+(\Delta_{c}+\vec{k}_c\cdot\vec{v})\sigma_{22}\nonumber\\
&+\Omega_{c}\sigma_{32}
+ga^{\dagger}\sigma_{13}+i\sqrt{\kappa_{1}}a^{\dagger}a_{p}\nonumber\\
&+\sum_{q}(g_{1q}\sigma_{31}e^{i(\omega_{p}-\vec{k}_a\cdot\vec{v}-\omega_q)t}+g_{2q}\sigma_{32}e^{i(\omega_{c}-\vec{k}_c\cdot\vec{v}-\omega_q)t})b_q\nonumber\\
&+\sum_{r}g_rd_r^\dagger ae^{i(\omega_r-\omega_{p}+\vec{k}_a\cdot\vec{v})t}+\textrm{h.c.},\label{eq:3}
\end{align}
where $\Delta_{a}=\omega_{a}-\omega_{p}$ represents the detuning of cavity mode and probe laser, $\Delta_{p}=\omega_{3}-\omega_{1}-\omega_{p}$ is the detuning of probe laser and the transition $\vert1\rangle\leftrightarrow\vert3\rangle$, and $\Delta_{c}=\omega_{3}-\omega_{2}-\omega_{c}$ is the detuning of the control laser and the transition $\vert2\rangle\leftrightarrow\vert3\rangle$.

\begin{figure}
\centering
\includegraphics[scale=0.25]{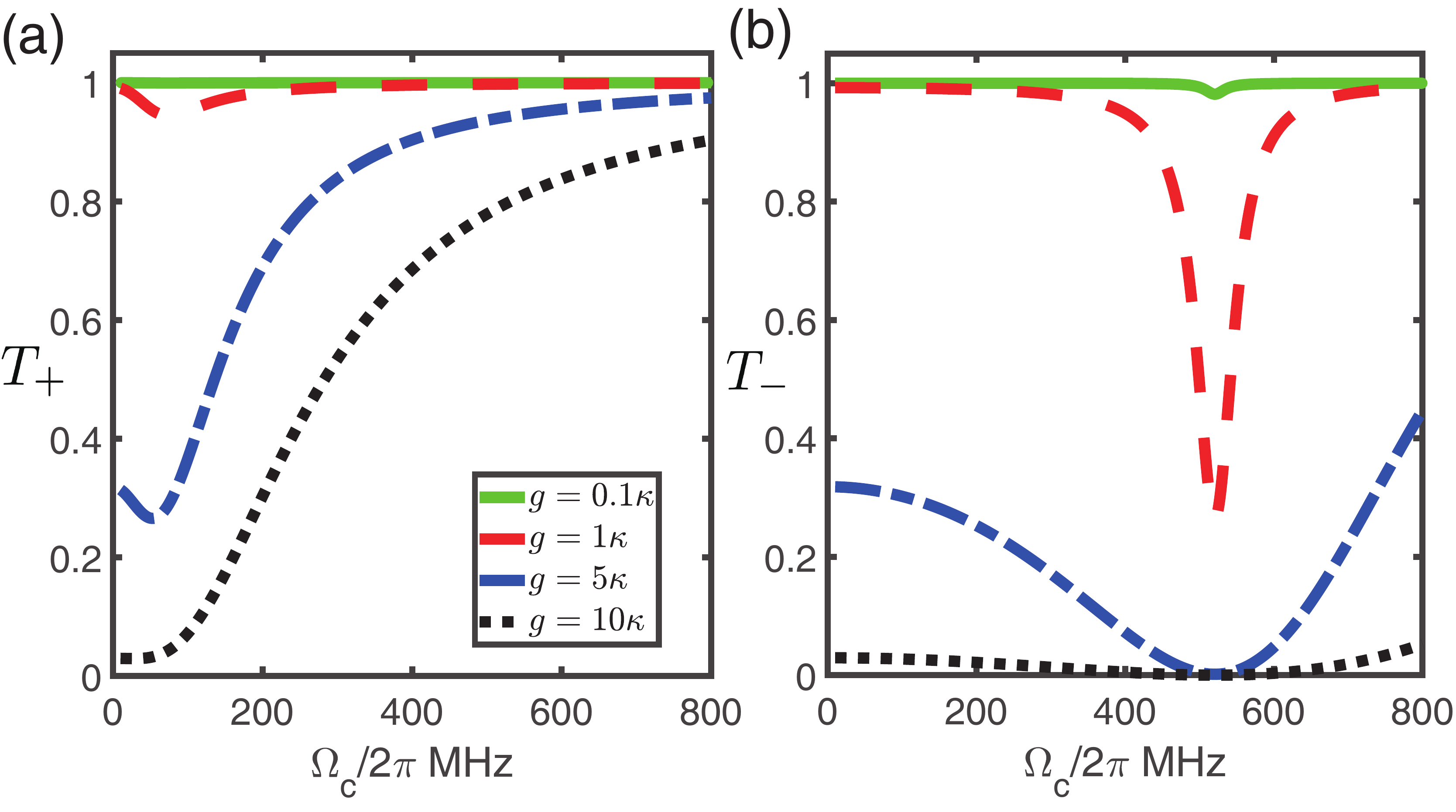}
\caption{The transmittance vs the Rabi frequency $\Omega_c$ of the control field for the probe light in the (a) co-propagation, and (b) counter-propagation case, respectively. The green solid line refers to the case with $g=0.1\kappa$, the red dashed line for $g=\kappa$, the blue dash-dotted line for $g=5\kappa$, and the black dotted line for $g=10\kappa$. The parameters are $v=250$~m/s, $\Delta_a=\Delta_p=0$, $\kappa_1=2\pi\times0.5$~MHz, $\kappa_2=2\pi\times4$~MHz, $\kappa_c=2\pi\times6$~MHz \cite{Zhang2018}, $\gamma_3=2\pi\times14.3$~MHz, and $\gamma_{12}=10.6$~MHz \cite{Zhou2017}.
\label{Fig2}}
\end{figure}

Among the methods for open quantum systems \cite{Gardiner1991,Walls1994,Tao2020}, the Heisenberg-Langevin approach can faithfully reproduce the quantum dynamics. Especially, it can significantly reduce the complexity of calculation as compared to the widely-used quantum master equation \cite{Breuer2007,Ai2014} and numerically-exact hierarchical equation of motion \cite{Ishizaki09-2,WangBX2018,Zhang2021}, when the system under investigation contains bosons and the number of operators of interest is small. In order to obtain the transmittance of the probe light, we apply the Heisenberg-Langevin approach to obtain %\cite{Wang2012}
\begin{eqnarray}
\dot{a}&=&-i(\Delta_{a}+\frac{\kappa}{2})a+\sqrt{\kappa_{1}}a_{p}-ig\sigma_{13}+F_a,\label{eq:4}\\
\dot{\sigma}_{13}&=&-[\gamma_{3}+i(\Delta_{p}+\vec{k}_a\cdot\vec{v})]\sigma_{13}-i\Omega_{c}\sigma_{12}\nonumber\\
&&+iga\left(\sigma_{33}-\sigma_{11}\right)+F_3,\label{eq:5}\\
\dot{\sigma}_{12}&=&-[\gamma_{12}-i(\Delta_{p}+\vec{k}_a\cdot\vec{v})+i(\Delta_{c}+\vec{k}_c\cdot\vec{v})]\sigma_{12}\nonumber\\
&&-i\Omega_{c}\sigma_{13}+iga\sigma_{32}+F_2,\label{eq:6}
\end{eqnarray}
where $\kappa=\kappa_1+\kappa_2+\kappa_c$, $\kappa_{1}$ ($\kappa_{2}$) is the coupling rate for the input (output) of the probe laser, and $\kappa_c$ is the intrinsic damping rate of cavity, $\gamma_3$ is the spontaneous decay rate associated with the electronic excited state $\vert3\rangle$, and $\gamma_{12}$ is the dephasing rate between the two ground states  $\vert1\rangle$ and $\vert2\rangle$. $F_a$, $F_3$, and $F_2$ are the Langevin noise operator of $a$, $\sigma_{13}$, and $\sigma_{12}$, respectively, which arise through the interaction with the reservoir, i.e.,
\begin{eqnarray}
%\kappa&=&2\pi g^2_r(\omega_p)D_2(\omega_p),\\
%\gamma_3&=&\pi[g^2_r(\omega_p-k_av)D_3(\omega_p-k_av)\nonumber\\
%&&+g^2_r(\omega_c-k_cv)D_2(\omega_c-k_cv)],\\
%\gamma_2&=&\pi g^2_r(\omega_c-k_cv)D_3(\omega_c-k_cv),\\
F_a&=&-i\sum_{r}g_rd_r(0)e^{-i(\omega_r-\omega_p)t},\\
F_3&=&i\sum_{q}[g_{1q}(\sigma_{33}-\sigma_{11})b_q(0)e^{-i(\omega_q-\omega_p+\vec{k}_a\cdot\vec{v})t}\nonumber\\
&&-g_{2q}\sigma_{12}b_q(0)e^{-i(\omega_q-\omega_c+\vec{k}_c\cdot\vec{v})t}],\\
F_2&=&i\sum_{q}[g_{1q}\sigma_{32}b_q(0)e^{-i(\omega_q-\omega_p+\vec{k}_a\cdot\vec{v})t}\nonumber\\
&&-g_{2q}\sigma_{13}b_q(0)e^{-i(\omega_q-\omega_c+\vec{k}_c\cdot\vec{v})t}].
\end{eqnarray}
%where $D_{3}$ ($D_{3}$) is the mode density of $d_r$ $(b_q)$.
Generally speaking, because the NV centers in nano-diamond interact with a complicated bath of phonons and spins, $\gamma_3$ and $\gamma_{12}$ show dependence on various factors, e.g. the temperature, the magnetic field, and the density of magnetic impurities \cite{Jarmola12}. By generalization of cluster correlation expansion, we can numerically simulate the open quantum dynamics of the NV center in the presence of spin bath at the low temperature \cite{Zhao2012,Yang2020}.

The initial state of NV center is prepared at $\vert1\rangle$ by optical pumping \cite{Zhou2017}. Next, we consider the steady-state solution by setting $\langle\dot{a}\rangle=0$, $\langle\dot{\sigma}_{13}\rangle=0$, and $\langle\dot{\sigma}_{12}\rangle=0$. In our configuration with $g\ll\Omega_c$, we have $\langle{\sigma}_{11}\rangle\approx1$ and $\langle{\sigma}_{32}\rangle\approx0$. Within the mean-field approximation, we have $\langle F_a\rangle=\langle F_3\rangle=\langle F_2\rangle=0$. By setting $\Delta_a=\Delta_c=0$ and solving Eqs.~(\ref{eq:4}-\ref{eq:6}), we can obtain
\begin{eqnarray}
\langle a\rangle_\pm&=&\frac{\sqrt{\kappa_1}\langle a_p\rangle}{i\Delta_p+\kappa/2+i\chi_\pm},\label{eq:10}\\
\chi_\pm&=&\frac{-i{\vert{g}\vert}^2}{\gamma_3+i(\Delta_p+k_av)+\frac{{\vert\Omega_c\vert}^2}{\gamma_{12}+i[\Delta_p+(k_a\mp k_c)v]}},\label{eq:11}
\end{eqnarray}
where $\chi$ is the susceptibility of the NV center to the cavity mode, the subscript $+$ ($-$) corresponds to the co-propagation (counter-propagation) case. Due to the mutual movement of NV center and light, the Doppler effect can not be ignored. In our system, the control laser is always injected from the left side, while the probe laser can propagate in either the same or the opposite direction. As shown in Fig.~\ref{Fig1}, we shall only consider the component of the velocity which is parallel to the direction of light propagation for the Doppler effect. Therefore, as shown in Fig.~\ref{Fig1}~(a) and (c), assuming that both the control and probe fields are of the same wave length, we have $\vec{k}_c\cdot\vec{v}=kv$ for both cases, while $\vec{k}_a\cdot\vec{v}=kv$ for the co-propagation and $\vec{k}_a\cdot\vec{v}=-kv$ for the counter-propagation respectively.

According to the input-output theory \cite{Gardiner1991,Walls1994}, the amplitude of the output field of the cavity is $\sqrt{\kappa_{2}}\langle a\rangle_+$ ($\sqrt{\kappa_{1}}\langle a\rangle_-$) in the co-propagation (counter-propagation) case.  Thus, the transmission spectra for the co-propagation (counter-propagation) case $T_{+}$ ($T_{-}$) can be written as \cite{Walls1994}
\begin{align}
T_{\pm}={\left\vert\frac{\sqrt{\kappa_1\kappa_2}}{i\Delta_p+\kappa/2+i\chi_{\pm}}\right\vert}^2.
\label{eq:T}
\end{align}
%with
%\begin{eqnarray}
%\chi_{\pm}&=&\frac{-i{\vert{g}\vert}^2}{\gamma_3+i(\Delta_p+kv)+\frac{{\vert\Omega_c\vert}^2}{\gamma_{12}+i\Delta_p}},\label{eq:13}\\
%\chi_{\textrm{cou}}&=&\frac{-i{\vert{g}\vert}^2}{\gamma_3+i(\Delta_p+kv)+\frac{{\vert\Omega_c\vert}^2}{\gamma_{12}+i(\Delta_p+2kv)}}.\label{eq:14}
%\end{eqnarray}
%\begin{figure*}[h]
%\centering
%\includegraphics[scale=0.36]{Fig3}
%\caption{The relationship between transmission, the control field strength $\Omega_c$ and the velocity $v$ with $\sqrt{N}g=0.1\kappa, \kappa, 5\kappa, 10\kappa$ respectively. (a)-(d) correspond to the co-propagation case and (e)-(h) correspond to the counter-propagation case. The red line in (g) is minimum value of transmission calculated analytically. The parameters are $\Delta_a=\Delta_p=0$~MHz, $\kappa_1=2\pi\times0.5$~MHz, $\kappa_2=2\pi\times4$~MHz, $\kappa_c=2\pi\times6$~MHz, $\gamma_3=2\pi\times10$~MHz, $\gamma_{12}=2\pi\times0.8$~MHz.
%\label{Fig3}}
%\end{figure*}
By the contrast of the two transmittance defined as \cite{Wang2013}
\begin{align}
\eta=\frac{T_{+}-T_{-}}{T_{+}+T_{-}},
\label{eq:C}
\end{align}
we can evaluate the nonreciprocity of the rotating NV center.

\section{Numerical Results and Analysis}
\label{sec:Results}

\begin{figure*}[htp]
\centering
\includegraphics[bb=90 0 1200 620,scale=0.45]{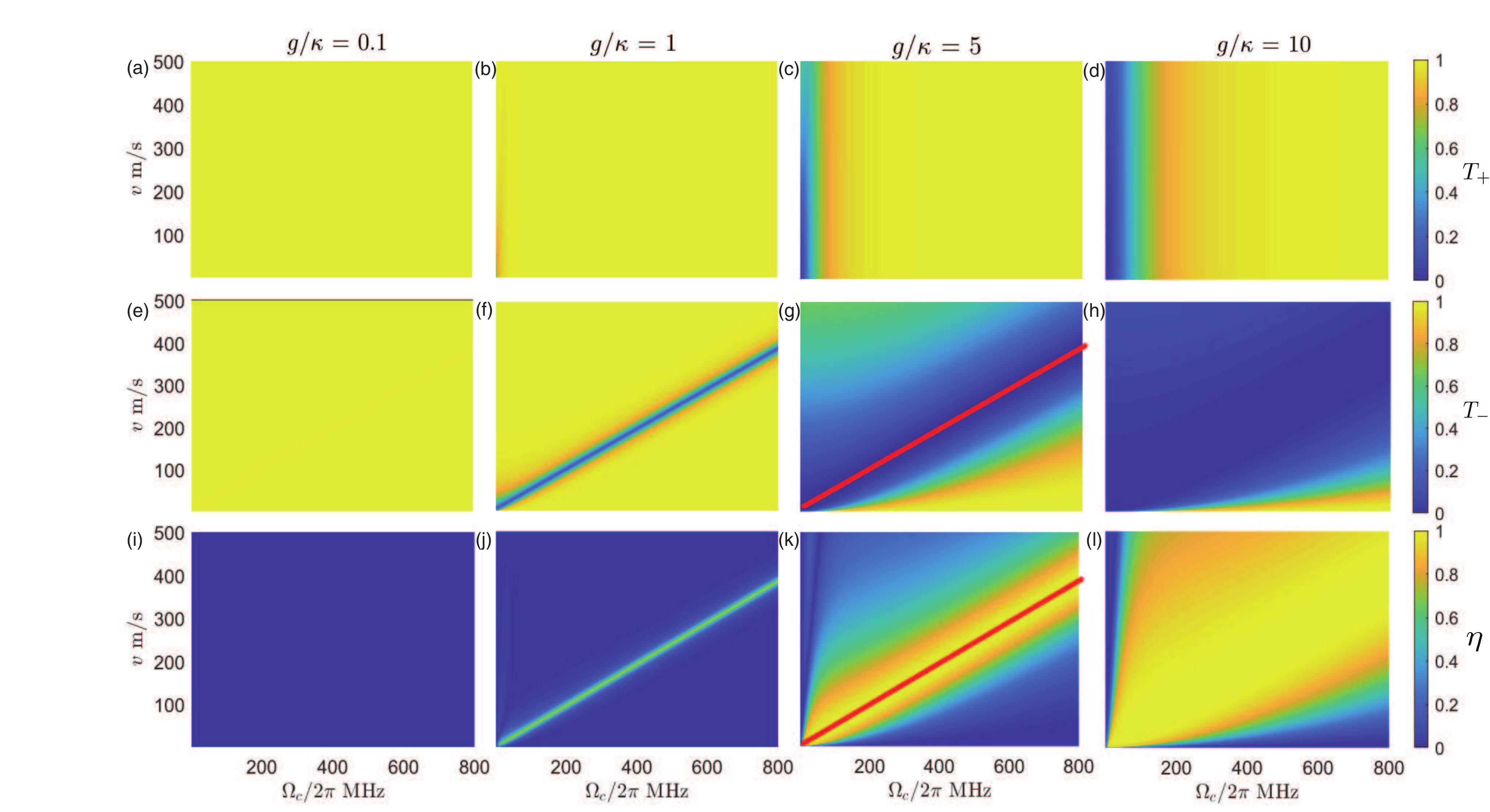}
\caption{The relationship between the transmittance, contrast, the control field strength $\Omega_c$ and the velocity $v$. The top (middle) row corresponds to the transmittance $T_+$ ($T_-$) in the co-propagation (counter-propagation) case. The bottom row shows the contrast $\eta$. From left to right, each column corresponds to $g/\kappa=0.1,~1,~5,~10$, respectively. The red lines in (g) and (k) is minimum value located at $kv=\sqrt{2}\Omega_c$.
\label{Fig3}}
\end{figure*}

We consider the nonreciprocal transport at the steady state. As shown in Fig.~\ref{Fig2}(a) for the co-propagation case with $v=250$~m/s, e.g. a nano-diamond rotating at 1~GHz with an NV center located at radius $0.25~\mu$m from the spin axis, we find a nearly-unity transport when $g\gg\kappa$ and $\Omega_c$ is large enough. This result demonstrates that the probe light explores the dark state for lossless transmission. In contrast, when the probe light propagates in the opposite direction of the control light, there is a fall around $\Omega_c=500$~MHz. This fall becomes more profound when $g$ is increased. Due to the term $2kv$ in $\chi_{-}$ of Eq.~(\ref{eq:11}), $\chi_{-}$ should achieve a maximum value at some $\Omega_c$.

In order to investigate the underlying physical mechanism for the nonreciprocal transport, we explore the dark-state mechanism in Ref.~\cite{WangYY2018}. The time evolution of the wave function is written as
\begin{widetext}
\begin{align}
\vert\psi(t)\rangle
 & =-\frac{\Omega_{c}}{\Omega}e^{-i\frac{g^{2}}{\Omega^{2}}\omega_{2}t}\left\vert E_{1}\right\rangle +\frac{g }{\sqrt{2}\Omega}e^{-\frac{i}{2}(\omega_{1}+\frac{\Omega_{c}^{2}}{\Omega^{2}}\omega_{2})t}(e^{-i\Omega t}\left\vert E_{2}\right\rangle +e^{i\Omega t}\left\vert E_{3}\right\rangle ),
\end{align}
where $\Omega=\sqrt{g^2+\Omega_{c}^2}$, $\left\vert E_{j}\right\rangle$'s are the three eigen states of the non-Hermitian system with $\left\vert E_{1}\right\rangle$ being the dark state and $\left\vert E_{2,3}\right\rangle$ being the bright states, $\omega_1=\delta-i\gamma_3$ with $\delta$ being single-photon detuning, $\omega_2=\Delta-i\gamma_{12}$ with $\Delta$ being two-photon detuning. The explicit expressions of the three eigen states are given as
\begin{align}
\left\vert E_{1}\right\rangle  & \simeq\frac{1}{N_{1}}[(\omega_{1}\omega_{2}-\Omega_{c}^{2})\left\vert1\right\rangle -g\omega_{2}\left\vert3\right\rangle +g\Omega_{c}\left\vert2\right\rangle ], \\
\left\vert E_{2}\right\rangle  & \simeq\frac{1}{N_{2}}\{[(\Omega-\omega_{1})(\Omega-\omega_{2})-\Omega_{c}^{2}]\left\vert1\right\rangle +g(\Omega-\omega_{2})\left\vert3\right\rangle +g\Omega_{c}\left\vert2\right\rangle \}, \\
\left\vert E_{3}\right\rangle  & \simeq\frac{1}{N_{3}}\{[(\Omega+\omega_{1})(\Omega+\omega_{2})-\Omega_{c}^{2}]\left\vert1\right\rangle -g(\Omega+\omega_{2})\left\vert3\right\rangle +g\Omega_{c}\left\vert2\right\rangle \},
\end{align}
with $N_i$'s being the normalization constants.
\end{widetext}

In our proposal, when the NV center is at rest, we set two-photon resonance for the control and probe fields, i.e., $\Delta=0$. When the NV center rotates along with the nano-diamond, the Doppler shifts will arise for the two fields. In the co-propagating case, since the Doppler shifts will be the same, the two-photon resonance still holds. When $g\ll\Omega_c$, cf. Fig.~\ref{Fig3}(a)-(b), the probe field is immune to absorption because the dark state dominates, i.e., $\Omega_c/\Omega\simeq1$, in the wave function, which decays at a lower rate of $(g^2/\Omega^2)\gamma_{12}$. If $g$ is increased, cf. Fig.~\ref{Fig3}(c)-(d), two factors will enhance the absorption. On the one hand, because $\omega_1\omega_2-\Omega_c^2\ll g\omega_{2},g\Omega_{c}$, $\vert2\rangle$ and $\vert3\rangle$ play a more significant role in $\vert E_1\rangle$. It makes the dark state more lossy, since it will decay at a larger rate as $g$ is increased. On the other, the two bright states will make a greater contribution, about $g/\sqrt{2}\Omega$, to $\vert\psi(t)\rangle$ with a faster decay rate $(\gamma_3+\kappa/2+\Omega_{c}^{2}\gamma_{12}/\Omega^{2})/2$. Therefore, we can observe a wider region for low transmittance as we increase $g$ in Fig.~\ref{Fig3}(a)-(d).

% and thus  \cite{WangYY2018}. However, in the counter-propagating case, because the Doppler shifts of the two fields will be of opposite sign, their detunings will become greater and greater as the velocity increases. Therefore, the probe field can not effectively transmit as the lossy component in the dark state becomes more significant \cite{WangYY2018}. Alternatively, during the evolution the dark state is subject to the absorption. In this way, the non-reciprocal transmission is realized.

%\begin{figure}[h]
%\centering
%\includegraphics[scale=0.19]{Fig4}
%\caption{The relationship between the contrast, the control field strength $\Omega_c$, and the velocity $v$. (a) to (d) corresponds to the case of $\sqrt{N}g=0.1\kappa, \kappa, 5\kappa, 10\kappa$ respectively. The yellow area indicates that the nonreciprocal transport is obvious. The red solid line in (c) represent the best position for nonreciprocal transport. The parameters are $\Delta_a=\Delta_p=0$~MHz, $\kappa_1=2\pi\times0.5$~MHz, $\kappa_2=2\pi\times4$~MHz, $\kappa_c=2\pi\times6$~MHz, $\gamma_3=2\pi\times10$~MHz, $\gamma_{12}=2\pi\times0.8$~MHz.
%\label{Fig4}}
%\end{figure}

\begin{figure}
\centering
\includegraphics[scale=0.25]{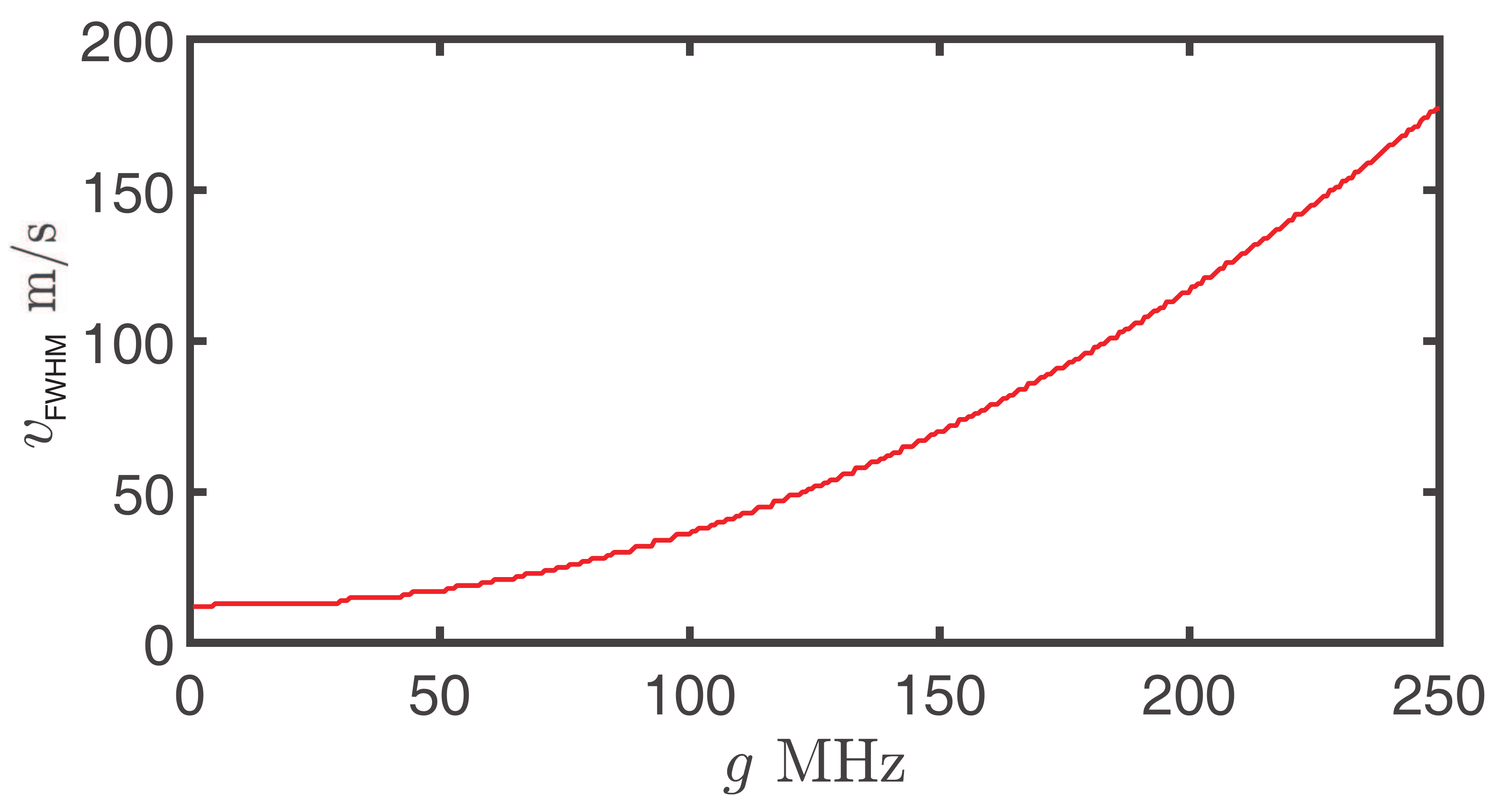}
\caption{The velocity $v_\textrm{FWHM}$ for the FWHM of $\eta$ vs $g$ in Fig.~\ref{Fig3}.
\label{Fig4}}
\end{figure}

If we turn to the counter-propagation case, we can observe a more interesting dependence of transmittance on $v$ and $\Omega_c$. When $g\ll\kappa$, as shown in Fig.~\ref{Fig3}(e), the transmittance is almost kept at unity for the whole parameter region. However, when $g=\kappa$ in Fig.~\ref{Fig3}(f), there emerges a dip in the transmittance for $kv=\sqrt{2}\Omega_c$. Along this line, we have $\omega_1\omega_2-\Omega_c^2\simeq2(kv)^2-\Omega_c^2=0$ and thus there is no component of $\vert 1\rangle$ in $\vert E_1\rangle$. The main contribution from $\vert E_2\rangle$ and $\vert E_3\rangle$ in $\vert \psi(t)\rangle$ results in the lossy in the transmission.
As $g$ increases, this area of exception becomes wider. Interestingly, the width of the dip is almost not influenced by either $v$ or $\Omega_c$, but $g$. According to our numerical simulation as shown in Fig.~\ref{Fig4}, the velocity corresponding to the full width at half maximum of the contrast is proportional to $g^2$.

\section{Conclusion and Discussion}
\label{sec:Conclusion}

In this paper, we explore the nonreciprocal transport in rotating nano-diamond induced by the EIT and the Doppler shift. We obtain the results at the steady state by the Heisenberg-Langevin approach. It is shown that the probe light makes use of the dark state and its transmission is generally not affected by the lossy intermediate state when it is incident in the same direction of the control light. However, the transmittance is significantly depressed around $kv=\sqrt{2}\Omega_c$ due to the breakdown of the two-photon resonance when the probe and control lights are in the opposite direction. Thus, we propose realizing optical nonreciprocity by using rotating nano-diamond with an NV center. We also discover that the velocity of the FWHM of contrast is proportional to $g^2$.

Previously, it was experimentally shown that the energies of the electronic ground states of NV centers in diamond will be shifted due to the rotation as a result of Barnett effect \cite{Barnett1915,Barnett1935,Wood2017,Chen2019}. However, since the energies of the electronic excited states will be correspondingly shifted by the same amount, the energy spectra in Fig.~\ref{Fig1} will not be modified by the rotation. Thus, the above proposed nonreciprocal transmission in rotating nano-diamond with an NV center still holds.

%When the coupling strength is weak, nonreciprocal transport can also be observed in a specific region. With the increase of the collective coupling strength, the region of nonreciprocal transport becomes wider.

We thank stimulating discussions with Jie-Qiao Liao, Jian Lin, and Yu-Qiao Li. This work is supported by the National Natural Science Foundation of China under Grant Nos.~11674033,~11474026,~11505007, and Beijing Natural Science Foundation under Grant No.~1202017.

\end{document}